\documentstyle[epsfig,psfig,sprocl]{article}
\begin{document}

\title{GAMMA-GAMMA, GAMMA-ELECTRON COLLIDERS: \\ 
  PHYSICS, LUMINOSITIES,  BACKGROUNDS.}

\author{ V.I. TELNOV }

\address{Institute of Nuclear Physics, 630090 Novosibirsk,
  Russia \\ email:telnov@inp.nsk.su }

\maketitle

\newcommand{\M}{\mbox{m}}
\newcommand{\n}{\mbox{$n_f$}}
\newcommand{\EE}{\mbox{ee}}
\newcommand{\EP}{\mbox{e$^+$}}
\newcommand{\EM}{\mbox{e$^-$}}
\newcommand{\EPEM}{\mbox{e$^+$e$^-$}}
\newcommand{\EMEM}{\mbox{e$^-$e$^-$}}
\newcommand{\GG}{\mbox{$\gamma\gamma$}}
\newcommand{\GE}{\mbox{$\gamma$e}}
\newcommand{\GP}{\mbox{$\gamma$e$^+$}}
\newcommand{\TEV}{\mbox{TeV}}
\newcommand{\GEV}{\mbox{GeV}}
\newcommand{\LGG}{\mbox{$L_{\gamma\gamma}$}}
\newcommand{\LGE}{\mbox{$L_{\gamma e}$}}
\newcommand{\LEE}{\mbox{$L_{ee}$}}
\newcommand{\WGG}{\mbox{$W_{\gamma\gamma}$}}
\newcommand{\EV}{\mbox{eV}}
\newcommand{\CM}{\mbox{cm}}
\newcommand{\MM}{\mbox{mm}}
\newcommand{\NM}{\mbox{nm}}
\newcommand{\MKM}{\mbox{$\mu$m}}
\newcommand{\SEC}{\mbox{s}}
\newcommand{\CMS}{\mbox{cm$^{-2}$s$^{-1}$}}
\newcommand{\MRAD}{\mbox{mrad}}
\newcommand{\IND}{\hspace*{\parindent}}
\newcommand{\E}{\mbox{$\epsilon$}}
\newcommand{\EN}{\mbox{$\epsilon_n$}}
\newcommand{\EI}{\mbox{$\epsilon_i$}}
\newcommand{\ENI}{\mbox{$\epsilon_{ni}$}}
\newcommand{\ENX}{\mbox{$\epsilon_{nx}$}}
\newcommand{\ENY}{\mbox{$\epsilon_{ny}$}}
\newcommand{\EX}{\mbox{$\epsilon_x$}}
\newcommand{\EY}{\mbox{$\epsilon_y$}}
\newcommand{\BI}{\mbox{$\beta_i$}}
\newcommand{\BX}{\mbox{$\beta_x$}}
\newcommand{\BY}{\mbox{$\beta_y$}}
\newcommand{\SX}{\mbox{$\sigma_x$}}
\newcommand{\SY}{\mbox{$\sigma_y$}}
\newcommand{\SZ}{\mbox{$\sigma_z$}}
\newcommand{\SI}{\mbox{$\sigma_i$}}
\newcommand{\SIP}{\mbox{$\sigma_i^{\prime}$}}

\abstracts {This report on Photon Colliders covers the following
  ``physics'' issues: physics motivation, possible luminosities,
  backgrounds, plans of works and international cooperation.
  More technical aspects such as  accelerator issues, new ideas
  on laser optics, laser cooling, and interaction region layout are
  discussed in my second talk at this Workshop.}

\section {Introduction}

In addition to \EPEM\ collisions, linear colliders provide a unique
possibility to study \GG\ and \GE\ interactions at energies and
luminosities comparable to those in \EPEM\ 
collisions.~\cite{GKST81}$^-$\cite{PToday}  High energy photons for
\GG, \GE\ collisions can be obtained using laser backscattering.
Modern laser technology presents the real possibility for construction
of the laser system for \GG, \GE\ collider ('photon collider').  This
option is now included in the pre-conceptual design of the NLC (North
American)~\cite{NLC}, TESLA (European)~\cite{TESLA} and JLC
(Asian)~\cite{JLC} linear collider projects in the energy range of a few
hundred GeV to about 1.5 TeV. These teams have intent so submit full
conceptual design reports in 2001-2002.
  However, in our time of tight HEP budgets the physics community
needs a very clear answer to the following question: a) can \GG,\GE\
collisions give new physics information in addition to \EPEM\
collisions that could justify an additional collider cost ($\sim$15\%,
including detector); b) is it technically feasible; c) is there enough
people who are ready to spend a significant part of their career
for the design and construction of a photon collider, and
exploiting its unique science?  

Shortly, my  answers are the following: 

a) Certainly yes. There are many predictions of extremely interesting
physics in the region of the next linear colliders. If something new will
be discovered (Higgs, supersymmetry or ... quantum gravity with extra
dimensions), to understand better the nature of these new
phenomena they should be studied in different reactions which give
complementary information.

b) There are no show-stoppers. There are good ideas on obtaining 
very high luminosities, on laser and optical schemes. It is clear how to
remove the disrupted beams and there is an understanding of backgrounds.
However, much remains to be done in terms of detailed studies and
experimental tests.  Special efforts are required for the development
of the laser and optics which are the key elements of photon
colliders.  

c) This is a new direction and it has to pass several natural phases
of development.  In the last almost two decade, a general conception
of photon colliders has been developed and has been discussed at many
workshops, the bibliography on \GG, \GE\ physics now numbers
over 1000 papers, mostly theoretical. The next phase will require much
wider participation of the experimental community.

 To this end, it was recently decided to initiate an International
collaboration on Photon Colliders.  This Collaboration does not
replace the regional working groups, but rather supports and
strengthens them. The Invitation letter, signed by Worldwide Study
contact persons on photon colliders: V.Telnov (Europe), K. Van Bibber
(North America), T.Takahashi (Asia) will be send to you shortly. 
\section{Physics}
\subsection{Higgs}
\begin{figure}[!htb]
 
\vspace{-1cm}
 
\hspace*{0cm}\begin{minipage}[b]{0.45\linewidth}
\centering
\vspace*{-0.cm}
\hspace*{-1.cm} \epsfig{file=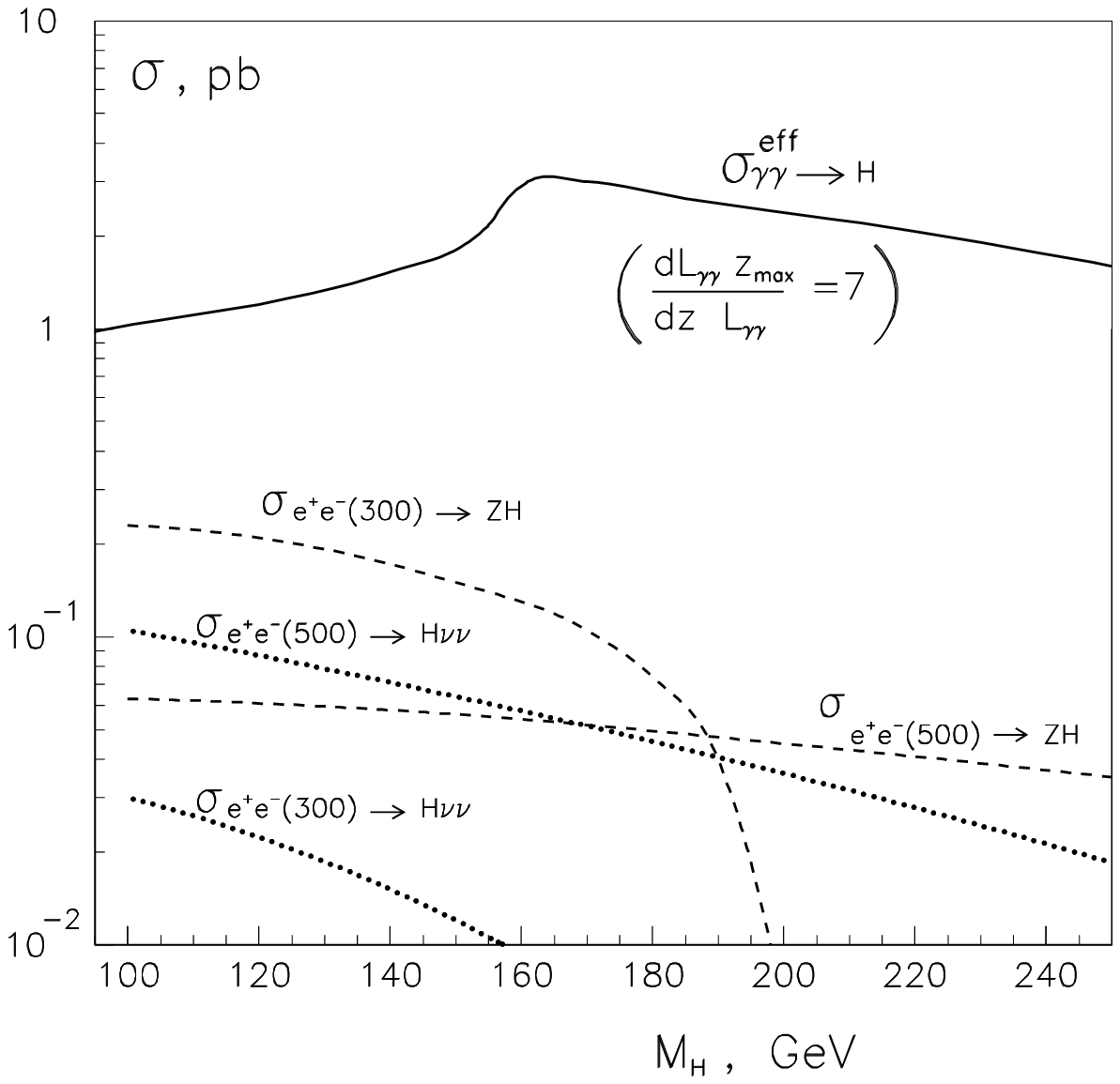,width=8.cm,angle=0}
 
\vspace{-0.4cm}
 
\caption{Cross sections for the Standard model Higgs in \GG\ and
 \EPEM\ collisions. \vspace{0.9cm}}
\label{cross}
\end{minipage}%
\hspace*{1.1cm} \begin{minipage}[b]{0.45\linewidth}
\centering
\vspace*{0.cm}

\hspace*{-1.4cm} \epsfig{file=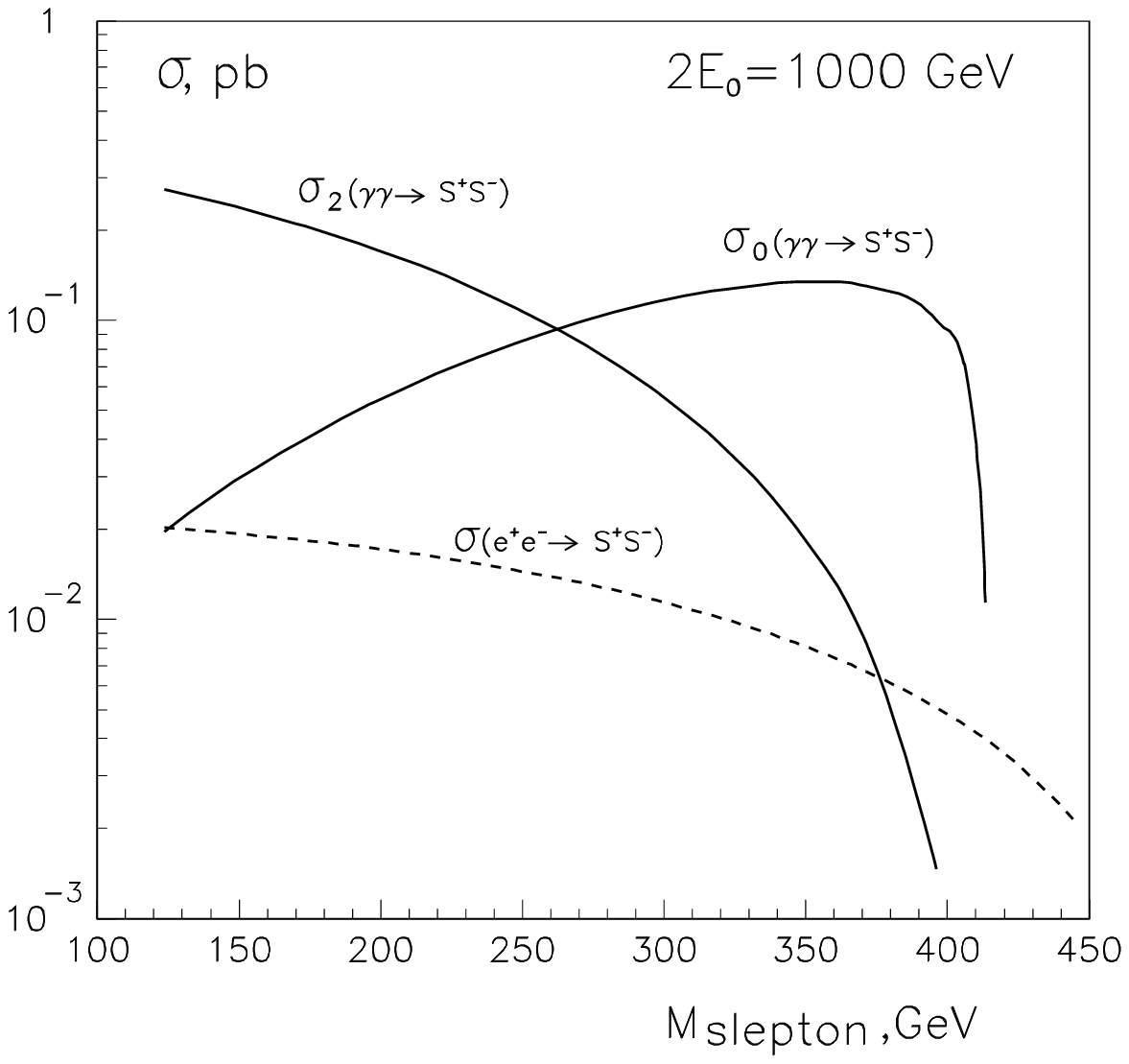,width=8cm,angle=0}
 
\vspace{-0.4cm}
 
\caption{Cross sections for charged scalars production in \EPEM\ and
\GG\ collisions at $2E_0$ = 1 TeV collider (in \GG\ collision
$W_{max}\approx 0.82$ TeV, $x=4.6$); $\sigma_0$ and $\sigma_2$ correspond to
the total \GG\ helicity 0 and 2.}
\label{crossel}
\end{minipage}
\end{figure}

The Higgs boson will be produced at photon colliders as a single
resonance.  This process goes via the loop and its cross section is
very sensitive to all heavy (even super-heavy) charged particles which
get their mass via the Higgs mechanism. The mass of the Higgs most
probably lies in the region of 100$<M_H<$250 GeV. The effective cross
section is presented in Fig.~\ref{cross}.~\cite{ee97}

Note that here \LGG\ is defined as the \GG\ luminosity at the high
energy luminosity peak ($z=\WGG/2E_e>0.65$ for $x=4.8$) with FWHM
about 15\%. For comparison, the cross sections of the Higgs production
in \EPEM\ collisions are shown.
We see that for $M_H=$ 120--250 GeV the effective cross section in
\GG\ collisions is larger than that in \EPEM\ collisions by a factor
of about 6--30.  If the Higgs is light enough, its width is much less
than the energy spread in \GG\ collisions. It can be detected as a
peak in the invariant mass distribution or can be searched for by energy
scanning using the very sharp ($\sim 1\%$) high energy edge of luminosity
distribution.~\cite{ee97}
The total number of events in the main decay channels $H \to b\bar b,
WW(W^*), ZZ(Z^*)$ will be several thousands for a typical integrated
luminosity of 10 fb.$^{-1}$ The scanning method also enables
the measurement of the Higgs mass with a high precision.
\subsection{Charge pair production}
   The second example is the charged pair production. It could be
   $W^+W^-$ or $t\bar{t}$ pairs or some new, for instance,
   supersymmetric particles.  Cross sections for the production of
   charged scalar, lepton, and top pairs in \GG\ collisions are larger
   than those in \EPEM\ collisions by a factor of approximately 5--10;
   for WW production this factor is even larger, about 10--20. The
   corresponding graphs can be found
   elsewhere.~\cite{TEL90}$^,$\cite{TESLA}$^,$\cite{ee97}

 The cross section of the scalar pair production
(sleptons, for example) in collision of polarized photons is shown
in Fig.\ref{crossel}.
One can see that for heavy scalars the cross section in collisions of
polarized photons is higher than that in \EPEM\ collisions by a factor
of 10--20. The cross section near the threshold is very sharp (in
\EPEM\ it contains a factor $\beta^3$) and can be used for
measurement of particle masses. 
Note that for scalar selectrons the cross section in \EPEM\ collisions
is not described by the curve in Fig.\ref{crossel} due to the existence of an
additional exchange diagram (exchange by neutralino). Correspondingly
the cross section is not described by pure QED (as it takes place in
\GG). Measurement of cross sections in both \EPEM\ and \GG\ channels
give, certainly, complementary information.
\subsection{Accessible masses}
 In \GE\ collisions, charged supersymmetric particles with masses
  higher than those in \EPEM\ collisions can be produced (a heavy charged
  particle plus a light neutral). \GG\ collisions also provide higher
  accessible masses for particles which are produced as a single
  resonance in \GG\ collisions (such as the Higgs boson). 

\subsection{Quantum gravity effects in Extra Dimensions.}
  This new theory~\cite{Arkani} is very interesting though beyond my
imagination. It suggests a possible explanation of why gravitation
forces are so weak in comparison with electroweak forces.
According to this theory the gravitational forces are as strong as
electroweak forces at small distances in space with extra dimensions
and became weak at large distances due to ``compactification'' of these
extra dimensions. 
  It turns out that this extravagant theory can be tested at linear
colliders and according to T.Rizzo~\cite{RIZZO} ($\GG\ \to WW$) and
K.Cheung~\cite{CHEUNG} ($\GG \to \GG$) photon colliders are sensitive
up to a factor of 2 higher quantum gravity mass scale than \EPEM\ collisions.
\section{Luminosity of photon colliders in current designs.}
\subsection{0.5--1 TeV colliders}
Some results of simulation of \GG\ collisions at TESLA, ILC (converged
NLC and JLC) and CLIC are presented below in Table~\ref{table1}. Beam
parameters were taken the same as those in \EPEM\ collisions with the
exception of the horizontal beta function at the IP which is taken (quite
conservatively) equal to 2 mm for all cases, that is several times
smaller than that in \EPEM\ collisions due to the absence of
beamstrahlung.  The conversion point(CP) is situated at distance
$b=\gamma\sigma_y$. It is assumed that electron beams have 85\%
longitudinal polarization and laser photons have 100\% circular
polarization.


\vspace{-0.3cm}
{\setlength{\tabcolsep}{2.mm}
{
\begin{table}[!hbtp]
\caption{Parameters of  \GG\ colliders based on Tesla(T), ILC(I)
and CLIC(C).}
\vspace{-0.0cm}
\begin{center}
\hspace*{-2.3mm}\begin{tabular}{c c c c c c c} \hline
 & T(500) & I(500) & C(500) \ &
  T(800) & I(1000) &  C(1000) 
                                                   \\ \hline \hline 
\multicolumn{7}{c}{ no deflection, $b=\gamma \sigma_y$, $x=4.6$} \\ \hline
$N/10^{10}$& 2. & 0.95 & 0.4 & 1.4 & 0.95 & 0.4 \\  
$\sigma_{z}$, mm& 0.4 & 0.12 & 0.05 & 0.3 & 0.12 & 0.05 \\  
$f_{rep}\times n_b$, kHz& 15 & 11.4 &30.1& 13.5 & 11.4 & 26.6 \\
$\gamma \epsilon_{x,y}/10^{-6}$,m$\cdot$rad & $10/0.03$ & $5/0.1$ & 
$1.9/0.1$&  $8/0.01$ & $5/0.1$ & $1.5/0.1$ \\
$\beta_{x,y}$,mm at IP& $2/0.4$ & $2/0.12$ & $2/0.1$ &
$2/0.3$& $2/0.16$ & $2/0.1$ \\
$\sigma_{x,y}$,nm& $200/5$ & $140/5$ & $88/4.5$ & 
$140/2$ & $100/4$ & $55/3.2$ \\  
b, mm & 2.4 & 2.4 & 2.2 & 1.5 & 4 & 3.1 \\
$L(geom),\,\,\,  10^{33}$& 48 & 12 & 10 & 75 & 20 & 19.5\\  
$\LGG (z>0.65), 10^{33} $ & 4.5 & 1.1 & 1.05 & 7.2 & 1.75 & 1.8 \\
$\LGE (z>0.65), 10^{33}$ & 6.6 & 2.6 & 2.8 & 8  & 4.2 & 4.6 \\
$\LEE, 10^{33}$ & 1.2 & 1.2 & 1.6 & 1.1 & 1.8 & 2.3 \\
$\theta_x/\theta{_y},_{max}$, mrad ~ & 5.8/6.5 & 6.5/6.9 & 6/7& 
 4.6/5 & 4.6/5.3 & 4.6/5.5 \\ \hline
\vspace{-5.mm}
\end{tabular}
\end{center}
\label{table1}
\end{table}
}} 

We see that \GG\ luminosity in the hard part of the spectrum $\LGG
(z>0.65)\sim 0.1L(geom)$, numerically it is about $(1/6)L_{\EPEM}$.
Note, that the coefficient $1/6$ is not a fundamental constant.  The
\GG\ luminosity in these projects is determined only by ``geometric''
ee-luminosity. With some new low emittance electron sources or with
laser cooling of electron beams after the damping ring (or photo-guns)
one can get, in principle, $\LGG (z>0.65) > L_{\EPEM}$. The
limitations and technical feasibility are discussed in the next
section and my second talk at this workshop.

Beside \GG\ collisions, there is considerable \GE\ luminosity (see
table) and it is possible to study \GE\ interactions simultaneously
with \GG\ collisions.
  
\begin{figure}[!htb]
\centering
\vspace*{-1.4cm} 
\hspace*{-1.4cm} \epsfig{file=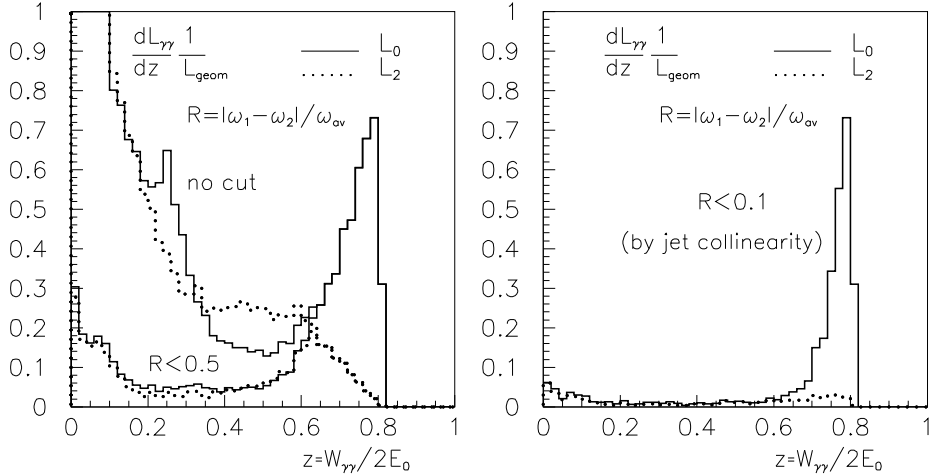,width=15.7cm,angle=0} 

\vspace{-1.6cm} 

\caption{\GG\ luminosity spectra at TESLA(500) for parameters
presented in Table 1. Solid line for total helicity of two photons 0
and dotted line for total helicity 2. Upper curves without cuts, two
lower pairs of curves  have cut on the relative difference
of the photon energy. See comments in the text.}

\vspace*{-2mm} 

\label{TeslaR}
\end{figure}

   The normalized \GG\ luminosity spectra for a 0.5 TeV TESLA are
   shown in Fig.\ref{TeslaR}(left).
The luminosity spectrum is decomposed into two parts, with the total
helicity of two photons 0 and 2. We see that in the high energy part
of the luminosity spectra photons have a high degree of polarization,
which is very important for many experiments.  In addition to the high
energy peak, there is a factor 5--8 larger low energy luminosity. It
is produced by photons after multiple Compton scattering and
beamstrahlung photons. Fortunately, these events have a large boost and
can be easily distinguished from the central high energy events.  In
the same Fig.\ref{TeslaR}(left) you can see the same spectrum with an
additional ``soft'' cut on the longitudinal momentum of the produced
system which suppresses low energy luminosity to a negligible level.

Fig.\ref{TeslaR} (right) shows the same spectrum with a stronger cut
on the longitudinal momentum. In this case, the spectrum has a nice
peak with FWHM about 7.5\%. On first sight such cut is somewhat
artificial because one can directly select events with high invariant
masses. The minimum width of the invariant mass distribution depends
only on the detector resolution. However, there is a very important
example when one can obtain a ``collider resolution'' somewhat better
than the ``detector resolution''; this is the
case of only two jets in the event when one can restrict the
longitudinal momentum of the produced system using the acollinearity
angle between jets ($H\to b\bar b, \tau\tau$, for example).

 A  similar table and distributions for the photon collider on the
c.m.s. energy 130 GeV (Higgs collider) can be found in
ref.\cite{TKEK}

\section{Ultimate \GG, \GE\  luminosities }
\begin{figure}[!htb]
\centering
\vspace*{-0.7cm} 
\hspace*{-0.8cm} \epsfig{file=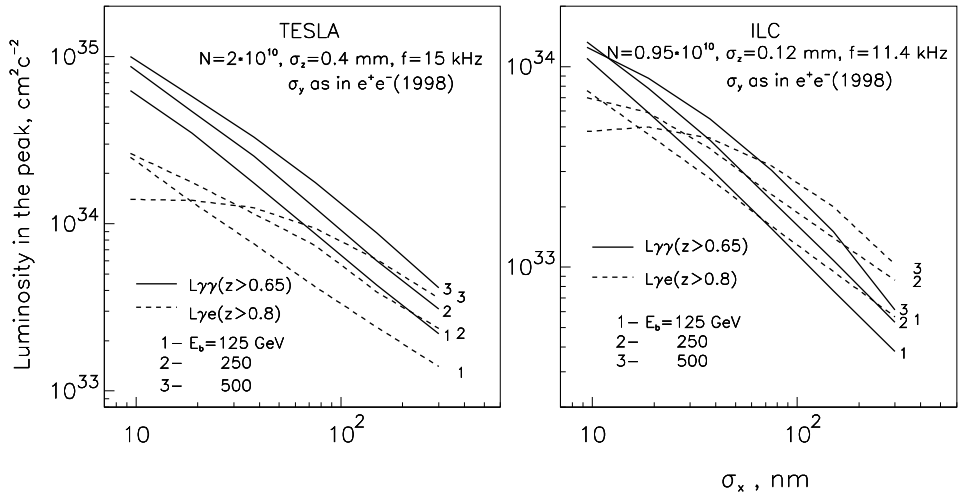,width=14.cm,angle=0} 
\vspace*{-1.5cm} 
\caption{Dependence of \GG\ and \GE\ luminosities in the high energy
peak on the horizontal beam size for TESLA and ILC at various
energies. See also comments in the text.}
\vspace{0mm}
\label{sigmax}
\vspace{1mm}
\end{figure}

The \GG\ luminosities in the current projects are determined by the
``geometric'' luminosity of the electron beams. Having electron beams
with smaller emittances one can obtain a much higher \GG\ 
luminosity.~\cite{TSB2} Fig.\ref{sigmax} shows dependence of the \GG\ 
(solid curves) and \GE\ (dashed curves) luminosities on the horizontal
beam size.
 The vertical emittance is taken as in TESLA(500), ILC(500)
projects (see Table \ref{table1}). The horizontal beam size was varied
by change of horizontal beam emittance keeping the horizontal beta
function at the IP constant and equal to 2 mm.

One can see that all curves for \GG\ luminosity follow their natural
behavior: $\L\propto 1/\sigma_x$, with the exception of ILC at
$2E_0=1$ GeV where at small $\sigma_x$ the effect of coherent pair
creation~\cite{CHEN}$^,$\cite{TEL90} is seen.\footnote{This curve has
  also some bend at large $\sigma_x$ that is connected with
  synchrotron radiation in quads (Oide effect) due to a large
  horizontal emittance. One can avoid this effect by taking larger
  $\beta_x$ and smaller \ENX.}  This means that at the same collider
the \GG\ luminosity can be increased by decreasing the horizontal beam
size at least by one order ($\sigma_x < 10$ nm is difficult due to
some effects connected with the crab crossing). Additional increase of
\GG\ luminosity by a factor about 3 (TESLA), 7(ILC) can be obtained by a
further decrease of the vertical emittance.~\cite{TKEK} So, using
beams with smaller emittances, the \GG\ luminosity at TELSA, ILC can
be increase by almost 2 orders of magnitude. However, even with one
order improvement, the number of ``interesting'' events (the Higgs,
charged pairs) at photon colliders will be larger than that in \EPEM\ 
collisions by about one order. This is a nice goal and motivation for
photon colliders.

  In \GE\ collision (Fig.\ref{sigmax}, dashed curves), the behavior of
the luminosity on $\sigma_x$ is different due to additional collision
effects: beams repulsion and beamstrahlung. As a result, the
luminosity in the high energy peak is not proportional to the
``geometric''  luminosity.

There are several ways of decreasing the transverse beam emittances (their
product): optimization of storage rings with long wigglers,
development of low-emittance RF or pulsed photo-guns with merging many
beams with low charge and emittances.  Here some progress is certainly
possible.  Moreover, there is one method which allows further decrease
of beam cross sections by two orders in comparison with current
designs. It is a laser cooling,~\cite{TSB1}$^,$\cite{Monter} see my
second talk at this workshop.

\section{Backgrounds}

Sometimes one hears that photon colliders are closer to pp than to
\EPEM\ colliders because the process $\GG\ \to hadron$ connected with
the hadronic component of the photon, which has a cross section by about 5
orders of magnitude larger than that of electromagnetic production of
charged pairs. Continuing this logics line one should say that \EPEM\ 
colliders are, in fact, rather photon colliders than \EPEM\ because the
cross section of the two-photon process $\EPEM\ \to\EPEM\ \EPEM\ $ is
10--11 orders higher than any of \EPEM\ annihilation processes. This
is obviously a misleading philosophy.

It is more correct to evaluate the seriousness of background by the problems
which it causes for experimentation: recording of data (trigger), their
analysis (underlying background processes, overlapping of interesting
and background events) and radiation damage of detector.  The proton
collider LHC has approximately the same luminosity as a photon
collider, but the hadronic background rate is 5 orders magnitude
higher; this causes radiation damage of detector components.  In this
respect photon colliders are much cleaner, practically the same as
\EPEM\ LC. Nevertheless, the background is a serious issue for both
\EPEM\ and \GG\ modes at an LC. This is connected mainly with the high
luminosity and relatively low beam collision rate that causes many
background reactions per each beam collisions.

  Let us enumerate the main sources of background at photon colliders:

  
$\bullet$ {\it Disrupted beams}. Low energy electrons after the multiple
  Compton scattering are deflected on opposing electron beam. The
  maximum disruption angle is about 10 mrad and the energy spread
 $ (0.02-1)E_0$. Solution: all these particles can be removed from
  the IP using the crab crossing collisions with $\alpha_c \sim 30$ mrad.
  
$\bullet$ {\it Electron-positron pairs}. This pairs are produced in the
  processes $\GG\ \to \EPEM,\; \gamma e \to \EPEM,\; \EE\ \to \EE\ + \EPEM$.
  There are unavoidable hard large angle particles with acceptable
  rate and many rather low $P_t$ electrons produced at very small angle
  and then kicked by the opposing electron beam. Due to solenoidal magnetic
  field these particles are confined in the region~\cite{TESLA} \\
$ r^2 [cm^2] < 0.12 (N/10^{10})(z[cm]/\sigma_z[mm]B[T]).$ \\
 The vacuum pipe should have larger radius. The level of \EPEM\ 
 background (mainly in the vertex detector) at photon colliders is is
 approximately the same as in \EPEM\ collisions though some additional
 study   taking into account ``reflection'' of
 particles from the mirrors is necessary .
 
$\bullet$ {\it Large angle Compton scattering}. The energy of these photon
is  $\omega = 4\omega_0/\theta^2$ at $\theta \gg 1/\gamma$, where
  $\omega_0$ is the energy of laser photons ($\sim$ 1 eV). 
At a distance $L$ the flux of photons   $dn/ds \propto
  N/\gamma^2L^2\theta^4$.  The main contribution comes from Compton scattering
  on low energy electrons.  The simulation for 2E = 500 GeV gives:
  $P \sim 10^{-7}$ W/cm$^2$, $\omega \sim 40$ keV at $\theta = 10$
  mrad (the edge of mirrors).

$\bullet$ {\it Large angle beamstrahlung}. The simulation shows that X-ray
  photons have a wide spectrum, $P \sim 10^{-6}$ W/cm$^2$, $\bar{\omega}
  \sim 1.5$ keV at $\theta = 10$ mrad. 
  
  X-rays may cause radiation damage problems for multilayer dielectric
  mirrors. For our case this problem is not sufficiently studied yet.
  In principle, there are dielectrical mirrors with very high
  radiation damage thresholds, sufficient for our task, it should be
  checked that they have simultaneously high reflectivity. In any
  case, one can use metal mirrors near the beam, for 1 $\MKM\ $ wave
  length the reflectivity is more than 99 \%.  Other problems with
  mirrors: change of the shape due to overheating and carbon deposits
  due to residual gas. Note, that the X-ray power density on the mirrors
  is proportional to $1/\vartheta^6$ and, if necessary, the minimum
  angle can be increased (it is very easy when the mirrors are place
  outside the beam).
  
$ \bullet$ {\it Halo of X-rays from final quads}. This is a problem
  for \EPEM\ colliders as well. The solution here is the scraping of
  electron beam tails by collimators before  final focusing. This is
  not a simple problem, especially if some halo arises after collimation.
 
 $ \bullet$ $\GG\ \to$ {\it hadrons}. Its cross section is $\sigma_{tot}
  \sim 500$ nb.$^{-1}$  For a typical case $\LGG \sim 10^{34},\;\nu
  \sim 10^4$ the background rate is 0.5 events/ bunch crossing.
  Hadronic background was studied in TESLA CDR.~\cite{TESLA} It will
  make problems for certain processes with jets at small angles (such
  as various QCD processes), however, for the ``main'' physics, where
  products usually have large angles, it  should be no serious
  problem even at maximum expected luminosities (one order higher
  than at the ``nominal'' TESLA).  It is important to develope
  algorithms of jet reconstruction which have low sensitivity to
  ``smooth'' hadronic background.  Influence of hadronic background on
  quality of reconstruction of various physics processes is one of
  important tasks of our Study.


%
\section{Conclusion}
   Prospects of photon colliders for particle physics are great; the physics
community should not miss this unique possibility.     

\end{document}